\documentclass[]{aa}
\usepackage{graphicx}
\usepackage[intlimits,sumlimits]{amsmath}
\usepackage{deluxetable}
\usepackage{times,epsfig} 
\usepackage{natbib}
\usepackage{amssymb}
\bibpunct{(}{)}{;}{a}{}{,} 
\usepackage{amsmath}
\usepackage[english]{babel}

\newcommand{\beqa}{\begin{eqnarray}} 
\newcommand{\eeqa}{\end{eqnarray}}
\newcommand{\Abst}[1]{\;#1}
\newcommand{\bsub}{\begin{subequations}}
\newcommand{\esub}{\end{subequations}}
\newcommand{\beal}{\begin{align}}
\newcommand{\ealn}{\end{align}}
\newcommand{\Nif}{$\rm ^{56}Ni$}

\newcommand{\ksm}{${\rm km~s^{-1}~Mpc^{-1}}$}

\begin{document}
\title{Consistent estimates of \Nif~yields for type Ia supernovae}
\titlerunning{SNe~Ia \Nif~mass estimates}
\authorrunning{Stritzinger et al.}
\author{M.~Stritzinger\inst{1,2}
 	\and P. A. Mazzali\inst{2,3} \and J. Sollerman\inst{1,4} \and S. Benetti\inst{5}}

\institute{Dark Cosmology Centre, Niels Bohr
Institute, University of Copenhagen, Juliane Maries Vej 30, DK-2100 
Copenhagen \O, Denmark\\
\email{max@dark-cosmology.dk}
\and  Max-Planck-Institut f\"ur Astrophysik, Karl-Schwarzschild-Str. 1,
85741 Garching bei M\"unchen, Germany 
\and  INAF-OATs, Via Tiepolo, 11, 34131 Trieste, Italy  \\
\email{mazzali@mpa-garching.mpg.de} 
\and Stockholm Observatory, AlbaNova, Department of Astronomy, 106 91 
Stockholm, Sweden \\
\email{jesper@dark-cosmology.dk}
\and INAF - Osservatorio Astronomico di Padova, vicolo
dell'Osservatorio 5, I-35122 Padova, Italy \\
\email{stefano.benetti@oapd.inaf.it}} 

\offprints{M. Stritzinger}
\date{Received -- / Accepted --}
\abstract{}{We present \Nif~mass estimates for seventeen well-observed 
type~Ia supernovae determined by two independent methods.}
{Estimates of the \Nif~mass for each type~Ia supernova are determined from
(1) modeling of the late-time nebular spectrum and (2) through
the combination of the peak bolometric luminosity with Arnett's rule.
The attractiveness of this
approach is that the comparison of estimated \Nif~masses circumvents errors
associated with the uncertainty in the adopted values of reddening and distance.
}
{We demonstrate that these two methods provide consistent 
estimates of the amount of \Nif~synthesized. 
We also find a strong correlation between the
derived \Nif~mass and the absolute $B$-band magnitude (M$_{B}$).}
{Spectral synthesis can be used as a diagnostic to study the 
explosion mechanism.
By obtaining more nebular spectra the \Nif--M$_{B}$ correlation
can be calibrated and can be used to investigate any 
potential systematic effects this relationship may have on the
determination of cosmological parameters, and provide a new way to 
estimate extra-galactic distances of nearby type Ia supernovae.}



\maketitle
\section{Introduction}

Type~Ia supernovae (hereafter SNe~Ia) are a vital 
tool for our understanding of the universe. Due to their high
luminosities and remarkable uniformity they are excellent cosmological distance 
indicators, and allow us to determine the evolution of cosmic expansion
\citep[see e.g.][$~$and references therein]{leibundgut01}.

SNe~Ia are used  as high 
precision cosmological probes \citep[see][]{riess04,astier06}. 
To maximize the level of precision achieved from current and future 
experiments requires the ability to understand and reduce any source of 
systematic error. The main motivation of this work was thus 
to learn more about the relationship between the \Nif~mass and
peak absolute magnitude.

To understand the dependence of the absolute brightness on
the \Nif~mass requires knowledge of the progenitor system 
and explosion mechanism of SNe~Ia.
Despite considerable efforts on the theoretical side to investigate the 
explosion physics, our current knowledge of the ignition mechanism and 
subsequent
flame propagation through the progenitor star is still rudimentary,
and in some instances debatable \citep{hillebrandt00}.
From an observational perspective, considerable effort has been put 
into supernova monitoring programs and many excellent data sets
have been assembled. With these data it is now possible to conduct detailed 
investigations 
into the physical mechanism leading to the observed diversity in the 
SNe~Ia population.  Some first steps have been made in this direction, e.g.
\citet{cappellaro97,mazzali98,contardo00,mazzali01,suntzeff03,benetti05,stritzinger06a,branch06}.

The amount of \Nif~produced in a SN~Ia explosion is a  key parameter that can 
be derived from observations and is important for our understanding of the 
luminosity decline-rate relation. 
In this article we demonstrate that the amount of \Nif~computed
from modeling the late-time nebular spectrum of a SN~Ia is consistent with
estimates obtained from the combination of the UltraViolet Optical
near-InfraRed (UVOIR) bolometric light curve with Arnett's rule
\citep{arnett82}.
We also find a strong correlation between the derived
\Nif~mass and M$_{B}$.  
As we can now estimate the mass of synthesized \Nif~
accurately from fitting the nebular spectrum this method may provide a 
means to understand possible sources of systematic error that could
effect the analysis of data from future high-z SNe~Ia surveys. 

The structure of this paper is as follows. In Sect.~2 we provide a brief 
description of the observational data considered in this study.
Section~3 describes the two methods used to estimate the \Nif~mass.
Section~4 contains the results, and is then followed by a discussion in 
Sect.~5.

\section{Observational data}

We have analyzed both published and unpublished photometric and spectroscopic CCD
data collected by numerous groups over the past two decades. A large
fraction of these observations were obtained by 
the Asiago group, the Center for Astrophysics group, and most 
recently by the European Supernova Collaboration. 
Data have also been taken from a variety of sources in the literature.
Two selection criteria were fulfilled by each SN~Ia in  
our sample: each event had to have comprehensive photometric $(U)BVRI$-band 
light curves that include maximum light as well as at least one nebular 
spectrum.
This spectrum is required for the spectral fitting method (see 
Sect. 3.1) while the well-sampled light curves are required to produce
a reliable UVOIR light curve (see Sect. 3.2). 

We were able to compile
nebular spectra and photometric data for seventeen events. 
Table~1 lists relevant information for each
event, which includes the  number of nebular spectra, adopted 
values of reddening and distance moduli, as well as an estimate of 
the $B$-band light curve decline rate parameter $\Delta$m$_{15}$($B$).

\section{Methods}
In this section we provide a short overview of the two methods used 
to determine the \Nif~mass for each SN~Ia.

\subsection{Modeling nebular spectra}
Here we summarize the spectral analysis method; for a more detailed description see \citet{mazzali97,mazzali98} and \citet{stehle05}.
A forthcoming publication will contain a detailed analysis 
of each modeled nebular spectra considered here (Mazzali, in preparation). 
This analysis includes constraints on nuclear
abundances, mixing within the ejecta, and limits on the extinction. 

The ratio between the explosion energy and the ejected mass in 
thermonuclear supernovae is larger compared to what is found in core collapse
supernovae.
As a result the ejecta in a SN~Ia become transparent to $\gamma$ rays
significantly earlier than in a type~II SN. After $\sim$150
days past maximum light the spectrum can safely be considered nebular. During
the nebular phase the emission spectrum is formed in the dense central regions
of the ejecta. It is at this location that the majority of Fe-group elements
are located.  As the abundance of
different Fe-group elements is a direct consequence of the nuclear burning
process, the modeling of this phase is an important
opportunity to constrain the explosion mechanism.

Synthetic spectra were calculated for each observed nebular spectrum  with a
model that uses a non-local thermodynamic equilibrium (NLTE) treatment of
the rate equations. The code first computes the deposition of the
$\gamma$ rays and positrons from the radioactive decay using a Monte Carlo
approach \citep{cappellaro97}. The deposition is transformed into collisional
heating following the prescriptions of Axelrod~~(1980; 
see also \citealt{rlplucy92}), and this is then
balanced by cooling via line emission. Both forbidden and allowed transitions
are included. Forbidden lines dominate the cooling, except for a few ions (Na~I,
Ca~II) where permitted lines dominate. Line emission is computed in concentric
shells within which the density and the abundances are constant. Since most
lines are blends, we observe the combined effect of density and abundance
distribution. The W7 model \citep{nomoto84} is used throughout for the density
distribution. A satisfactory fit with observed line profiles is made by 
changing the abundances in each shell.
Rather than trying to fit the broadest features, which are formed by the blend
of many weak lines, we instead fit the two strongest Fe-group features
Fe~III  at 4650~\AA\ and Fe~II + Fe~III at  5300~\AA. 
This gives a more accurate estimate of the line emissivity and hence of the
\Nif~mass. 

\begin{figure}
\resizebox{\hsize}{!}{\includegraphics{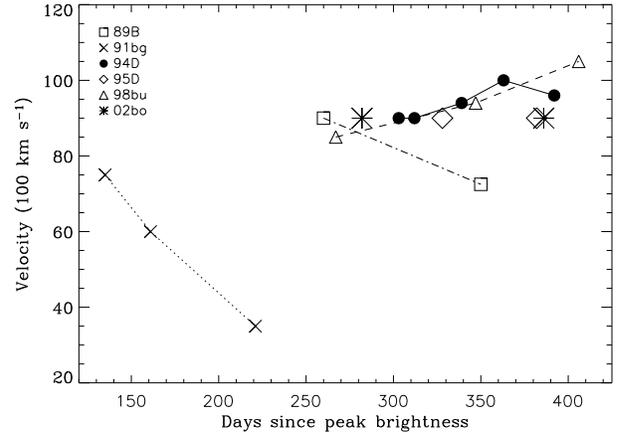}}
\caption{The evolution of the modeled velocity for six SNe~Ia as a function of
time past maximum brightness.}
\end{figure}

Based on two infrared nebular spectra, \citet{spyromilio04} reported evidence 
for an increase of the expansion velocities as a function of time. 
We also find an increase in the emission line velocities of SN~1998bu.
However, with the total data set considered in this study, which comprises 
six SNe~Ia with at least two spectra (see Table~1),
we cannot confirm an increase in line velocities.
This is shown in Fig. 1 where we plot the velocity used in the model for the
Fe nebula vs.  days past maximum brightness. For expansion velocities 
greater than $\sim$6000 km~s$^{-1}$ the 
modeled nebular velocity is linearly correlated with the FWHM velocity of 
the two Fe lines that we fit \citep[see][]{mazzali98}.
For all normal SNe~Ia we find some dispersion ($\sim$10\%) about 
a mean value, however
there is no clear trend for any evolution in the line velocities. 
Only SN~1991bg shows a significant decrease in line velocities. 
This may be due to the ejecta not having
reached the complete nebular phase when the first two spectra were obtained.
Moreover, it is known that severe blending of lines occurs between
velocities of 3000 to 6000 km s$^{-1}$ making it difficult to obtain a 
very reliable fit. 
For events with more than one nebular spectrum we computed the \Nif~mass 
for each spectrum and then averaged the estimates. 

The main uncertainty affecting our estimate of the \Nif~mass 
(excluding uncertainties in reddening and distance) comes from
the unknown contribution in the infrared (IR), which is known to increase
during late-times \citep{sollerman04}. Some species, in particular Si, 
emit only in the IR, and their abundance cannot be constrained directly by modeling
the optical spectrum only. However, in the best observed IR spectrum of a SN~Ia
in the nebular phase (SN~1998bu, \citeauthor{meikle98}, priv. comm.;  
Spyromilio et al. 2004) Si emission lines do not make a major contribution. 
Most of the IR flux is emitted from Fe and Co lines, and much of it 
is therefore taken into account in our models. 
We estimate that the error in our estimate of
the \Nif~mass including errors associated with our code, the flux contribution 
in the IR, and possible systematic errors due to uncertain
calibration of the late-time spectra, is of the order of $\sim$15\% (see below).


\subsection{The UVOIR light curve and Arnett's rule}

A complete description of how to derive the \Nif~mass using
broad-band optical photometry is presented in \citet{contardo00}. 
The overall strategy is to construct a 
quasi-bolometric light curve which allows us to avoid the complicated
radiative transfer required to interpret single passband light curves. 
The early-time UVOIR light curve of a SN~Ia is known to include most of the
total {\em true} bolometric flux \citep{suntzeff96,contardo00}. 
At maximum light the UVOIR light curve does not account for the $\sim$5\% 
contribution of flux emitted in the IR nor any flux blue-wards of 3200~\AA. 
These contributions add up to less than $\sim$15\% of the total flux. 

Following \citeauthor{contardo00}, the $(U)BVRI$ photometric light curves
are fitted with a 10-parameter function. This function contains no physics,
rather it produces a continuous description of each light curve and a set 
of parameters that can be compared to one another.
To calculate a UVOIR light curve each filter light curve is summed 
and a correction is made to account for the extinction. 
When $U$-band data is lacking (see Table~1) a correction based on SN~1992A 
\citep{suntzeff96} is applied in the manner described by 
\citet{contardo00}. The absolute flux scale
is set with an accurate distance\footnote{When possible a direct distance 
measurement, e.g. a Cepheid distance estimate, is adopted.
If no direct distance measurement is available we used relative 
distances to Virgo \citep{kraankorteweg86} adopting a Virgo distance
of 15.3 Mpc or distances from the Hubble flow.}. 

To derive the \Nif~mass we make use of Arnett's rule \citep{arnett82},
which states that at the time of maximum light the SN~Ia luminosity is
equal to the energy inputs from the radioactive decays within the
expanding ejecta.
A derivation of an empirical relation for Arnett's rule that connects 
the observed luminosity at maximum light to the \Nif~mass is given in 
\citet{stritzinger05}. Assuming a rise time to bolometric maximum of 19 days,
this simple relation gives a total luminosity at maximum light of

\beal
\label{nimass}
{\rm L_{max}}~=~2.0~\times~10^{43}~\left(\frac{\rm M_{Ni}}{\rm M_{\sun}}\right)~~{\rm erg~s^{-1}}
\Abst{.}
\end{align}

\noindent To account for flux outside the optical and near-IR passbands a 
10\% correction is added to each \Nif~mass estimate obtained from Eq.~(1).

Recently, we conducted an investigation \citep[][]{blinnikov06} of Arnett's 
rule using synthetic light curves computed using the 1-D hydro code STELLA
\citep{sorokina03}. This showed that with the UVOIR light curve and
this 10\% correction, Arnett's rule estimates the \Nif~yield accurately to
with in $\lesssim$0.05 M$_{\sun}$.
\begin{figure}
\resizebox{\hsize}{!}{\includegraphics{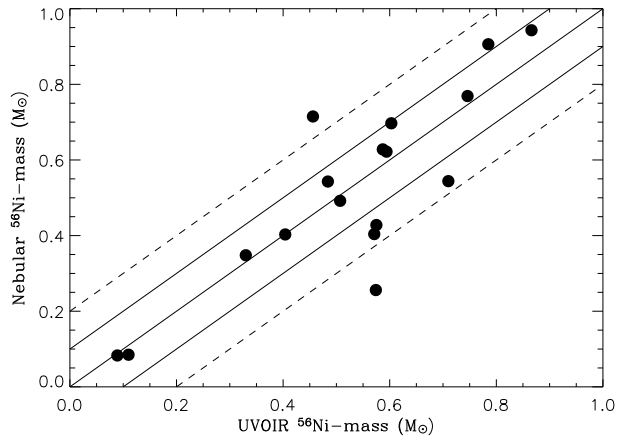}}
\caption{Plot of \Nif~mass derived for seventeen SNe~Ia
via the late-time nebular spectrum fitting method vs. the
UVOIR bolometric light curve method. Dashed and dotted diagonal lines
correspond to 10\% and 20\% offsets from the slope of the solid line,
respectively.}
\label{niplot}
\end{figure}

\section{Results}

We now estimate the \Nif~mass for seventeen SNe~Ia. Figure~\ref{niplot} 
contains the \Nif~mass determined from the synthetic spectral analysis of the
nebular spectra plotted vs. the \Nif~mass determined from the UVOIR light 
curves. These values are listed in Table~2. 
Because we compare estimates of the \Nif~mass derived from two independent
methods, our results are not affected by the uncertainties associated with
the adopted reddening and distance values.  We have therefore not included 
error bars arising from the uncertainity in these quantities.
Including the uncertainties in distance and extinction for each event
can lead to an uncertainty in the absolute value of each \Nif~mass 
estimate from $\sim$0.05 to $\sim$0.35~M$_{\sun}$
\citep{stritzinger06a}.
Instead, in Fig.~2 are plotted offsets from the line with a slope of one.
These correspond
to 10\% (dashed lines) and 20\% (dotted lines) respectively.
From Fig.~\ref{niplot} it is evident that eleven of the seventeen
SNe~Ia have \Nif~masses that agree with each other to within 10\%,
four to within 20\%, and only two differ by more than 20\%.

\begin{figure}
\resizebox{\hsize}{!}{\includegraphics{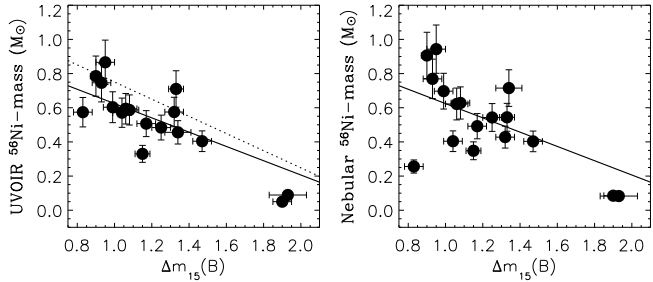}}
\caption{\Nif~mass derived from the UVOIR light curve method
({\em left}) and the nebular spectrum fitting method
({\em right}) vs. $\Delta$m$_{15}$($B$). The solid line in both panels
correspond to the Phillips M$_{B}$--$\Delta$m$_{15}$($B$)
relation \citep{phillips99}. Dotted lines in the left panel corresponds
to the Phillips M$_{B}$--$\Delta$m$_{15}$($B$) relation when the adopted
bolometric correction  of $-0.28$ is changed to $-0.48$.}
\label{ni.m15B}
\end{figure}

In order to assess the intrinsic uncertainty of the spectral synthesis 
method we varied the size of the error bars on the \Nif~mass estimates
and computed the reduced $\chi^{2}$. 
It is found that an uncertainity of 20\% clearly overestimates 
the true random uncertainty while 10\% gives a reduced $\chi^{2} = 4.0$.
We find that an uncertainty between 15\% and 20\% gives reasonable
reduced $\chi^{2}$ values. For error bars that are 15\% of the \Nif~mass
estimates we obtain a reduced $\chi^{2} = 1.8$. Furthermore if we 
include a 3-$\sigma$ clipping algorithm, which excludes one event
(in this cases SN~1995al) we obtain a reduced $\chi^{2} = 1.16$. Clearly
the error bar for SN~1995al should be larger than 15\%. 
Moreover, events with poor quality data such as SN~1989B, which
has two low signal-to-noise spectra and suffers from high values of
extinction may also have an uncertainity larger than 15\%.
Nevertheless in the following we conclude that 
an average 15\% uncertainty for the error bars is reasonable for the
adopted uncertainties in the values of the 
\Nif~mass estimates from the spectral synthesis method 
determined with good quality data. As discussed above the \Nif~mass estimates
determined via the UVOIR light curve and Arnett's rule are accurate to within
$\sim$10\%. To be conservative we also adopt a 15\% uncertainty for this method.
\begin{figure}
\resizebox{\hsize}{!}{\includegraphics{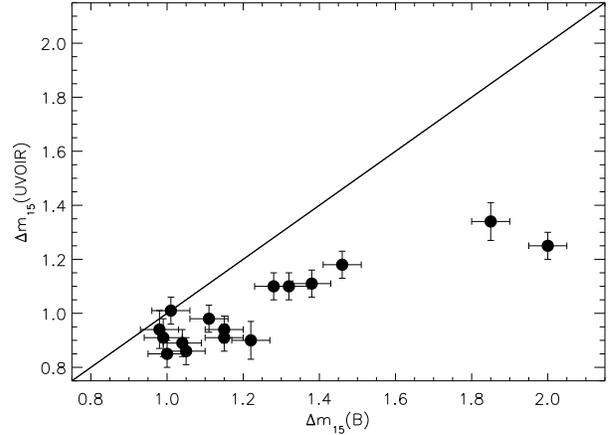}}
\caption{$\Delta$m$_{15}$($UVOIR$) vs. $\Delta$m$_{15}$($B$). Solid line
has a slope of one. We assume a $\Delta$m$_{15}$($UVOIR$) uncertainty of
0.05 for each event except those that have a $U$-band
correction. In those cases the uncertainity was increased to 0.07.}
\label{uvoir.m15}
\end{figure}

In Fig.~3 the \Nif~mass derived for each event via both methods is plotted vs. 
$\Delta$m$_{15}$($B$). 
The solid line in both panels corresponds to the 
M$_{B}$--$\Delta$m$_{15}$($B$) relation given in Phillips et al. (1999,
see their Table~3). To convert M$_{B}$ to a UVOIR
luminosity we used the bolometric correction (BC) given by 
\citet{branch92} of $-$0.28, and then Eq.~1 to convert luminosity to a 
\Nif~mass. To illustrate the effect of the adopted BC on the Phillips relation
we have included in the left panel of Fig.~3 (dotted line) the 
M$_{B}$--$\Delta$m$_{15}$($B$) relation if the BC is decreased to
a rather conservative value of $-0.42$. As shown in Fig.~3 this leads to a 
steepening of the Phillips relation.
Clearly the \Nif~mass estimates from both methods are correlated to 
$\Delta$m$_{15}$($B$). Moreover, this correlation 
is in excellent agreement with the the Phillips M$_{B}$--$\Delta$m$_{15}$($B$)
relation. Note that a least squares fit computed from
the data gives a line that is very similar to the Phillips relation line. 
In both cases the least squares line is only slightly steeper.

We found no  correlation between \Nif~mass and $\Delta$m$_{15}$($UVOIR$) 
for events with $\Delta$m$_{15}$($UVOIR$) $\leq$ 1.1. 
This motivated us to plot in Fig.~4 $\Delta$m$_{15}$($UVOIR$) vs. 
$\Delta$m$_{15}$($B$).\footnote{By plotting the decline rate parameters rather
than a parameter(s) set by an absolute flux scale, we 
bypass the effects associated with the uncertainty in the adopted distance.}
The solid line corresponds to a slope of one. For all events
$\Delta$m$_{15}$($UVOIR$) is less than $\Delta$m$_{15}$($B$) except for
SN~2000cx which has identical values for both parameters.
Evidently $\Delta$m$_{15}$($UVOIR$) is correlated with
$\Delta$m$_{15}$($B$). Note there are events
with similar values of $\Delta$m$_{15}$($UVOIR$) that show a scatter in 
$\Delta$m$_{15}$($B$) of $\sim$0.4 mag.
Finally, the fact that the estimates of $\Delta$m$_{15}$($B$) and 
$\Delta$m$_{15}$($UVOIR$) are identical for SN~2000cx is 
most likely due to its asymmetric rise (and fall) to (from)
maximum light \citep[see][]{li01}.

\section{Discussion} 

The agreement between the \Nif~mass estimates shown in Fig.~2 is
encouraging as these two methods are independent and use different 
observational data.
The ability to accurately probe the explosion mechanism
with one spectrum taken more than 100 days after the explosion
reveals the power of the spectral synthesis method. 

The range of a factor of ten in the amount of 
synthesized \Nif, as indicated by the spectral synthesis method, is consistent 
with results obtained from previous studies of UVOIR light curves 
\citep{suntzeff96,cappellaro97,contardo00,strolger02,suntzeff03,stritzinger06a}. 
The observed range in \Nif~mass is difficult to reconcile with the current SN~Ia
paradigm \citep{nomoto84,woosley86,hillebrandt00}, 
i.e. the thermonuclear disruption of a Chandrasekhar-size white dwarf. 
This has lead to the introduction of the so-called delayed detonation
model \citep{khokhlov91,woosley90,woosley94a,hoflich96}, and most recently
more exotic explosion mechanisms \citep{bravo06}.
An alternative to the Chandrasekhar mass model that 
addresses the observed diversity of SNe~Ia properties is one involving 
a variable progenitor mass.
There is observational evidence that sub-Chandrasekhar mass white
dwarfs are a plausible candidate for the less luminous events and at
least for the sub-luminous SN~1991bg-like
variety of SNe~Ia \citep{mazzali97,cappellaro97,stritzinger06a}. 
Another possible progenitor maybe a super-Chandrasekhar mass progenitor
\citep{yoon04,yoon05}.

Efforts to model the light curves of SNe~Ia led to the suggestion
that the absolute luminosity at maximum light [hence
$\Delta$m$_{15}$($B$)] is a direct consequence of the 
\Nif~mass \citep{arnett82,hoeflich96,cappellaro97,mazzali01,pinto01,kasen06}.
The strong correlation shown in Fig.~3 between the
\Nif~mass and $\Delta$m$_{15}$($B$) provides observational evidence 
that supports this hypothesis. 
This argument is strengthened by
the excellent agreement between the Phillips M$_{B}$--$\Delta$m$_{15}$($B$)
relation and the results in Fig.~3. With the acquisition of more nebular 
spectra it should be possible to calibrate the \Nif~mass--M$_{B}$
relation. With this calibration one could then use a nebular spectrum to 
(1) determine the distance to a nearby SN~Ia, and more importantly (2) 
address the relationship between the \Nif~mass and the peak absolute magnitudes.
This method is independent of the luminosity-decline rate relation, so 
distances determined through this method are not as susceptible to the
secondary parameters that may affect the peak phase light curve
\citep{hamuy96,mazzali98,tripp99,hantano00,benetti04,benetti05}. 
The late-time bolometric luminosity, however, may be susceptible to 
the distribution of isotopes and/or the fraction of positron kinetic energy 
that is deposited within the ejecta.
Nonetheless, by gaining a deeper understanding of how 
the peak absolute brightness relates to the \Nif~mass we increase our 
understanding of the explosion mechanism and any  
possible systematic errors incurred via this relationship. 

The scatter seen between $\Delta$m$_{15}$($UVOIR$)
and $\Delta$m$_{15}$($B$) hints at the complexities associated with the 
transfer of radiation through expanding ejecta, and how time-dependent
NLTE effects influence the flux evolution of the near-IR light curves. 
Indeed a comparison between the near-IR light curves for events
with similar values of $\Delta$m$_{15}$($B$) indicates a diversity in their 
morphology. 

The knowledge afforded by the spectral synthesis method is significant
and warrants further SNe~Ia observational programs to
obtain well-calibrated nebular spectra. 

\begin{acknowledgements}
The Dark Cosmology Centre is funded by the Danish National Research Foundation. 
M.S. and P. M. are grateful to Wolfgang Hillebrandt for his generous
hospitality. This work is supported in part by the European 
Community's Human Potential Programme under contract HPRN-CT-2002-00303, ``The
Physics of Type Ia Supernovae''.
This research has made use of the NASA/IPAC 
Extragalactic Database (NED), which is operated by the Jet Propulsion 
Laboratory, California Institute of Technology, under contract with the 
National Aeronautics and Space Administration.

\end{acknowledgements}


\clearpage
\begin{deluxetable}{l c c r c c c c}
\tablecolumns{8}
\tablenum{1}
\tablewidth{0pc}
\tablecaption{The SNe~Ia sample}
\label{data.tab}
\tablehead{
\colhead{SN} &
\colhead{Ref.$^{a}$} &
\colhead{N$^{b}$} &
\colhead{Filters} &
\colhead{Ref.$^{c}$} &
\colhead{E($B-V$)$_{\rm tot}^{d}$} &
\colhead{$\mu^{e}$} &
\colhead{$\Delta$m$_{15}$($B$)$^{f}$}}

\startdata
1989B  &  W94  & 2 & $BVRI$  & W94  & 0.370 &  29.86  &  1.34$\pm$0.07 \\
1990N  &  AA   & 1 & $UBVRI$ & L98  & 0.020 &  31.90  &  1.08$\pm$0.05 \\
1991T  &  AA   & 1 & $UBVRI$ & L98  & 0.140 &  30.74  &  0.95$\pm$0.05 \\
1991bg &  T96  & 3 & $BVRI$  & F92, L93, T96  & 0.060 &  31.13  &  1.93$\pm$0.10 \\
1992A  &  M97  & 1 & $UBVRI$ & S96            & 0.020 &  31.41  &  1.47$\pm$0.05 \\
1994D  &  M97  & 5 & $UBVRI$ & R95, P96, M96, S00 & 0.040 &  30.90  &  1.32$\pm$0.05 \\
1994ae &  M97  & 1 & $BVRI$  & R99                & 0.150 &  32.22  &  0.90$\pm$0.03 \\
1995D  &  M97  & 2 & $BVRI$  & R99                & 0.090 &  32.43  &  0.99$\pm$0.05 \\
1995al &  AA   & 1 & $BVRI$  & R99                & 0.160 &  32.00  &  0.83$\pm$0.05 \\ 
1996X  &  S01  & 1 & $UBVRI$ & R99,S01            & 0.060 &  32.02  &  1.25$\pm$0.05 \\
1998bu &  C01  & 3 & $UBVRI$ & J02, S99           & 0.350 &  29.97  &  1.04$\pm$0.05 \\
1999by &  J06    & 1 & $UBVRI$ & G04                & 0.000 &  30.03  &  1.90$\pm$0.05 \\
2000cx &  S04  & 1 & $UBVRI$ & J02, Li01, C03     & 0.080 &  32.64  &  0.93$\pm$0.05 \\
2001el &  M05 & 1 & $UBVRI$ & K03                & 0.240 &  30.55  &  1.15$\pm$0.04 \\
2002bo &  B04  & 2 & $UBVRI$ & B04                & 0.380 &  31.67  &  1.17$\pm$0.05 \\
2002er &  K05  & 1 & $UBVRI$ & P04                & 0.360 &  32.90  &  1.33$\pm$0.04 \\ 
2003du &  VS06 & 1 & $UBVRI$ & VS06               & 0.000 &  32.75  &  1.06$\pm$0.06 \\

\enddata
\tablenotetext{a}{Reference to the nebular spectrum.}

\tablenotetext{b}{N is the number of nebular spectra used to estimate the 
\Nif~mass.}

\tablenotetext{c}{Reference(s) to the photometric data.}
\tablenotetext{d}{Reddening values are from the references quoted in 
column 5 or are at least consistent with their estimate.}
\tablenotetext{e}{Cepheid distances for SNe~1989B, 1991T, and 1998bu
taken from \citet{freedman01} 
(H$_{\circ}$ $=$ $72$ \ksm). For the remaining
events, the relative distances to Virgo are taken from 
\citet{kraankorteweg86} adopting a Virgo distance of 15.3 Mpc or distances
from the Hubble flow.}

\tablenotetext{f}{If available, values of $\Delta$m$_{15}$($B$) were taken 
from \citet{benetti05}, otherwise were taken from \citet{phillips99} or
from the literature.}

\tablerefs{ 
(AA)   Asiago SN archive;
(B04)  \citealt{benetti04};
(C01)  \citealt{cappellaro01};
(C03)  \citealt{candia03};
(F92)  \citealt{filippenko92};
(G04)  \citealt{garnavich04};
(J02)  \citealt{jha02};
(J06)  \citealt{jha06};
(K05)  \citealt{kotak05};
(K03)  \citealt{krisciunas03};
(L93)  \citealt{leibundgut93};
(L98)  \citealt{lira98};
(L01)  \citealt{li01};
(M96)  \citealt{meikle96}; 
(M97)  \citealt{mazzali97};
(M05)  \citealt{mattila05};
(P96)  \citealt{patat96};  
(P04)  \citealt{pignata04};
(R95)  \citealt{richmond95};
(R99)  \citealt{riess99};
(S00)  \citeauthor[][$~$priv. comm.]{smith00}; 
(S01)  \citealt{salvo01};
(S04)  \citealt{sollerman04}; 
(S96)  \citealt{suntzeff96};
(S99)  \citealt{suntzeff99};
(T96)  \citealt{turatto96}; 
(VS06) \citealt{stanishev06};
(W94) \citealt{wells94}.}


\end{deluxetable}

\clearpage
\begin{deluxetable}{c c c c}
\tablecolumns{4}
\tablenum{2}
\tablewidth{0pc}
\tablecaption{Derived SNe~Ia parameters}
\label{data.tab}
\tablehead{
\colhead{$\Delta$m$_{15}$($UVOIR$)} &
\colhead{\Nif($UVOIR$)} &
\colhead{\Nif(nebular)} &
\colhead{SN}\\
\colhead{} &
\colhead{(M$_{\sun}$)} &
\colhead{(M$_{\sun}$)} &
\colhead{} 
} 

\startdata
 0.85  &  0.87 & 0.94 & 1991T \\
 0.85  &  0.57 & 0.26 & 1995al \\
 0.86  &  0.59 & 0.63 & 1990N \\ 
 0.89  &  0.59 & 0.62 & 2003du \\
 0.90  &  0.46 & 0.72 & 1989B \\
 0.91  &  0.33 & 0.35 & 2001el \\
 0.94  &  0.60 & 0.70 & 1995D  \\
 0.94  &  0.51 & 0.49 & 2002bo \\
 0.95  &  0.79 & 0.91 & 1994ae \\
 0.98  &  0.57 & 0.40 & 1998bu \\
 1.02  &  0.75 & 0.77 & 2000cx \\
 1.10  &  0.48 & 0.54 & 1996X \\
 1.10  &  0.71 & 0.54 & 2002er \\
 1.11  &  0.40 & 0.40 & 1992A \\
 1.18  &  0.58 & 0.43 & 1994D \\
 1.25  &  0.05 & 0.09 & 1999by \\
 1.42  &  0.09 & 0.08 & 1991bg \\
\enddata

\end{deluxetable}

\end{document}